\documentclass[a4paper]{jpconf}
\usepackage{graphicx}
\usepackage{cancel}
\begin{document}
\title{Physics of the Interplay Between the Top Quark and the Higgs
  Boson\footnote{Talk given by J. Santiago at Top 2012, Winchester,
    UK, September 16-21, 2012.}}

\author{Mikael Chala and Jos\'e Santiago}

\address{CAFPE and Departamento de F\'{\i}sica Te\'orica y del Cosmos,
  \\ University of Granada, 18071, Granada, Spain}

\ead{miki@ugr.es, jsantiago@ugr.es}

\begin{abstract}
We discuss some aspects of the interplay between the top quark and the
Higgs boson at the LHC. First we describe what 
indirect information on the top Yukawa coupling can be extracted 
from measurements in the Higgs sector. We then show that the study of
processes involving $Ht\bar{t}$ and $Hb\bar{b}$ final states can give
us information on the spectrum in models of strong electroweak
symmetry breaking.
In particular, we introduce a
novel analysis of the $Hb\bar{b}\to bb\bar{b}\bar{b}$ channel with an
excellent discovery potential at the LHC.
\end{abstract}

\section{Introduction}

With the discovery of the Higgs boson at the LHC, the quest to
understand the mechanism of electroweak symmetry breaking (EWSB) has finally
started. The top quark has an undeniable leading role in the physics
of the Higgs boson, as it dominates, in the Standard Model (SM), the main
production mechanism ($gg\to H$) and contributes in an important way
to one of the most relevant decay channels ($H\to \gamma
\gamma$). Hence, experimental tests of the Higgs sector can give us
precious information on aspects of the top quark, like its Yukawa
coupling or on the possible existence of top partners or even new
vector resonances. At the same
time the LHC is a top factory and many of the top quark properties
will be measured to an unprecedented accuracy. Searches of processes
involving the top quark or some of its possible partners can in turn
give us non-trivial information on the sector of EWSB. 
In these proceedings we will discuss some aspects of this
interplay between the top quark and the Higgs boson. In
Section~\ref{whathiggstells} we will describe what indirect
information on the top quark Yukawa coupling we can infer from
current results on Higgs searches and how that information might be
combined with other searches to obtain a coherent picture of the
EWSB sector. 
In Section~\ref{newproduction} we show how searches involving
the production of $Ht\bar{t}$ and $Hb\bar{b}$ can give us information
on the spectrum in models of strong EWSB. In these models, new
vector-like quarks can be singly produced
in association with a SM quark, mediated by a color octet
resonance. Subsequent decays of the heavy quark can result in $H
t\bar{t}$ and $H b \bar{b}$ final states. We show that these channels
can be measured at the LHC for a large region of parameter space. In
particular we introduce a novel analysis of the $Hb\bar{b}\to
bb\bar{b}\bar{b}$ channel that takes full advantage of the distinctive
kinematics of the process, leading to an excellent reach potential at
the LHC.
We finally
conclude in Section~\ref{conclusions}.

\section{What is the Higgs telling about the top?\label{whathiggstells}}

Now that the discovery of a Higgs-like boson has been settled all the
efforts are directed towards measuring its main properties. Even
though the errors are still large, a coherent picture is starting to
emerge from the combination of all the
different channels~\cite{Higgs:combination,:2012zzl}. The experimental data
point to a new boson of mass
$m_H\approx 126$ GeV with production and decay compatible with those of
the standard model Higgs in all channels with the possible exception
of the $H\to \gamma \gamma$ decay, which seems to be consistently
larger than the SM one (but still compatible with it at the two
standard deviation
level).~\footnote{Currently ATLAS also observes an incompatibility between
  the mass of the new boson as measured in the $ZZ^\ast$ and $\gamma
  \gamma$ channels at the level of almost three standard deviations.}

The $H\to \gamma \gamma$ decay is unique for two reasons. First, it
occurs at the loop level in the SM, being \textit{a
  priori} particularly sensitive to new physics. Second, this decay arises from
interfering contributions with the top and the $W$ boson running in
the loop and it is therefore sensitive not only to the
absolute value of the top Yukawa coupling but also to its sign.
It turns out that although it is still possible to explain
the data with new particles contributing at the loop level, it
would not be easy to accommodate the current enhancement 
without modifying the production cross section
through gluon fusion and without destabilizing the vacuum (see for
instance~\cite{Carena:2012xa}).
If we assume instead that 
there is no significant contribution to the Higgs production and
decay from new particles and the only effect is a possible
modification of the couplings of the SM particles to the Higgs we find
that apart from a region with SM-like couplings there is a second
possible explanation to the current data. In this region the couplings
of the gauge boson to the Higgs are slightly
reduced whereas the top Yukawa coupling has the opposite
sign than in the SM~\cite{fits}. 
The ``wrong sign'' top Yukawa coupling turns the destructive interference
in the $H\to \gamma \gamma$ decay into a constructive one thus
explaining the observed enhancement.

Most physical processes are sensitive to the absolute value of the top
Yukawa and not to its sign so it is not easy to test the hypothesis of
a wrong sign Yukawa coupling. One possibility, recently proposed
in~\cite{Farina:2012xp}, is to use t-channel
single top production in association with a Higgs
boson. This is a tree level
process in which there is interference between diagrams involving
the top Yukawa and a $W$, making it also
sensitive to the sign of the top Yukawa coupling relative to the $WWH$
coupling. The process is very
small in the SM due to an almost perfect destructive
interference. With the wrong sign top Yukawa coupling however it
becomes observable at the LHC and it is therefore a potentially
crucial test of this explanation of current Higgs data. It should be
noted however that the information on the top Yukawa coupling is
obtained from a loop process in one case and from a tree level process
in the other. These two Yukawa couplings are in general different as
new physics can affect differently the tree and loop processes. Therefore
care should be exercised when interpreting the results of these searches.

\section{New Higgs Production Mechanisms in Composite Higgs
  Models\label{newproduction}}

\begin{figure}[t]
\begin{minipage}{0.45\textwidth}
\vspace{0.1cm}
\includegraphics[width=0.95\textwidth]{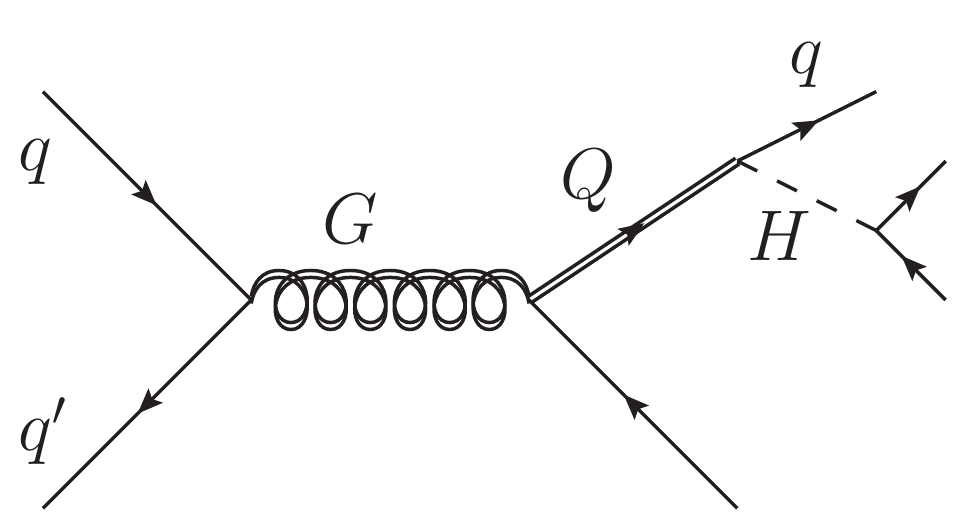}
\vspace{1cm}
\caption{Main diagram contributing to $Ht\bar{t}$ and $Hb\bar{b}$
  production in our model. Here $Q$ ($q$) stands for $T$ ($t$) or $B$
  ($b$).
\label{diagram}
}
\end{minipage}
\hfil
\begin{minipage}{0.45\textwidth}
\includegraphics[width=0.95\textwidth]{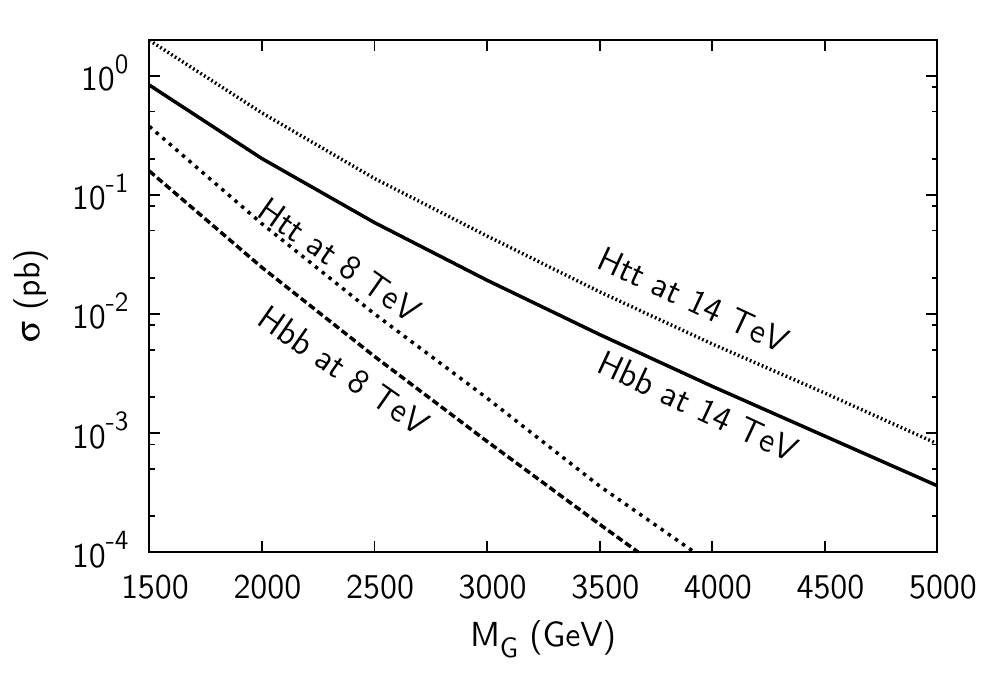}
\caption{$Ht\bar{t}$ and $Hb\bar{b}$ 
production cross section as a function of $M_G$ for 
$\sin (\phi_{t_R}) = \sin (\phi_{b_R})=0.6$.
\label{Hqq:xsec}}
\end{minipage}
\end{figure}

We have seen in the previous section how Higgs searches can give us
information on some properties of the top quark like its Yukawa
coupling. In this section we will see how searches involving the
production of $H t\bar{t}$ and $H b\bar{b}$ can give us information on the
spectrum of new particles in models of strong EWSB.
Many models of strong EWSB include in their spectrum light top
partners (new vector-like quarks that mix strongly with the top and/or
bottom quarks) together with massive color-octet vector resonances, 
that we will call heavy gluons. In some
regions of parameter space the leading production mechanism of these
top partners is single production mediated by the exchange of the
heavy gluons. If the top partners have electric charge $2/3$ or $-1/3$
they can result in $Ht\bar{t}$ and
$Hb\bar{b}$ final states, respectively. The corresponding process is
shown in Fig.~\ref{diagram}. 
In the following we discuss
how to use these channels to search for the top partners and even for
the heavy gluons in these models of strong EWSB. 
In our studies we have used a simplified version of the minimal
composite Higgs model~\cite{Agashe:2004rs} as described
in~\cite{Bini:2011zb,Carmona:2012jk} (see also~\cite{Barcelo:2011wu}).
The most relevant parameter
that we will consider is the degree of compositeness
of the top or bottom quarks, denoted by $\sin (\phi_{t_R})$ and $\sin
(\phi_{b_R})$, respectively. The $Ht\bar{t}$ and $Hb\bar{b}$ production
cross sections, for $\sin (\phi_{t_R}) = \sin (\phi_{b_R})=0.6$ are
reproduced in Fig.~\ref{Hqq:xsec} at the LHC for $\sqrt{s}=8$ and $14$ TeV.
We have simulated our signal and backgrounds using
\texttt{MADGRAPH V4.5.0}~\cite{Alwall:2007st} and \texttt{ALPGEN
  V.2.13}~\cite{Mangano:2002ea}, respectively. We have then passed the
events through
\texttt{PYTHIA 6.4}~\cite{Sjostrand:2006za} for hadronization and
showering and \texttt{DELPHES 
  V1.9}~\cite{Ovyn:2009tx} for detector simulation. Further details on
the implementation of the model and the simulations can be found in
reference~\cite{Carmona:2012jk}.

\subsection{$H t\bar{t}$ Channel}

We consider the process shown in Fig.~\ref{diagram} with $Q=T$ and
$q=t$ and the leading $H\to b\bar{b}$ decay
channel of the Higgs and a semileptonic decay of the $t\bar{t}$
system. The specific process is
\begin{equation}
pp\to G \to T\bar{t}+\bar{T}t \to H t\bar{t} \to 4b+ 2j+ l \cancel{E}_T.
\end{equation}
A detailed analysis of this channel 
has been performed in Ref.~\cite{Carmona:2012jk} where all the
relevant details can be found. Here we will just summarize the main
results for the LHC running at a center of mass energy $\sqrt{s}=14$
TeV. The heavy gluon masses that can be probed at this energy are
large enough to make all the decay products quite hard. Also $T$ is
typically heavy so that the use of boosted techniques probed useful in
these searches. 

In order to take maximum advantages of the particular
features of the signal we have implemented the following set of cuts
\begin{itemize}
\item At least 3 jets, with a minimum of 2 b tags.
\item At least 1 isolated charged lepton.
\item The two jets with the largest invariant masses, $j_{1,2}$, 
  are required to have invariant masses close to
  the top and Higgs mass, respectively, $|m_{j_1}-m_t|\leq 40$ GeV and
  $|m_{j_2}-m_H|\leq 40$ GeV.
\item A cut on $S_T$ (the scalar sum of the $p_T$ of the three hardest
  jets, the charged lepton and the missing transverse energy)
that depends on the test $M_G$ we are considering
\begin{equation}
S_T>1.2,~1.5,~1.7,~2\mbox{ TeV for }M_G=2,~2.5,~3,\geq 3.5\mbox{ TeV}.
\end{equation}
\end{itemize}
We have considered all the relevant backgrounds in our analysis. The
most important ones turn out to be $t\bar{t}$ and $t\bar{t} b\bar{b}$.
The result of this analysis is summarized in Fig.~\ref{Htt:results} in
which we show the sensitivity that can be reached, for different
values of $M_G$, with an integrated luminosity of 100 fb$^{-1}$ as a
function of $\sin (\phi_{t_R})$.

\subsection{$Hb\bar{b}$ Channel}

We now turn our attention to the case $Q=B$ and $q=b$. 
We will show that contrary to naive expectations,
the $H\to b\bar{b}$ decay channel, resulting in a four b quark final
state, can be quite competitive, allowing for the reconstruction not
only of the heavy quark but also of the heavy gluon. The process we
are interested in is
\begin{equation}
pp\to G \to B\bar{b}+\bar{B}b \to H b\bar{b} \to 4b.
\end{equation}

\begin{figure}[t]
\begin{minipage}{0.45\textwidth}
\includegraphics[width=0.95\textwidth]{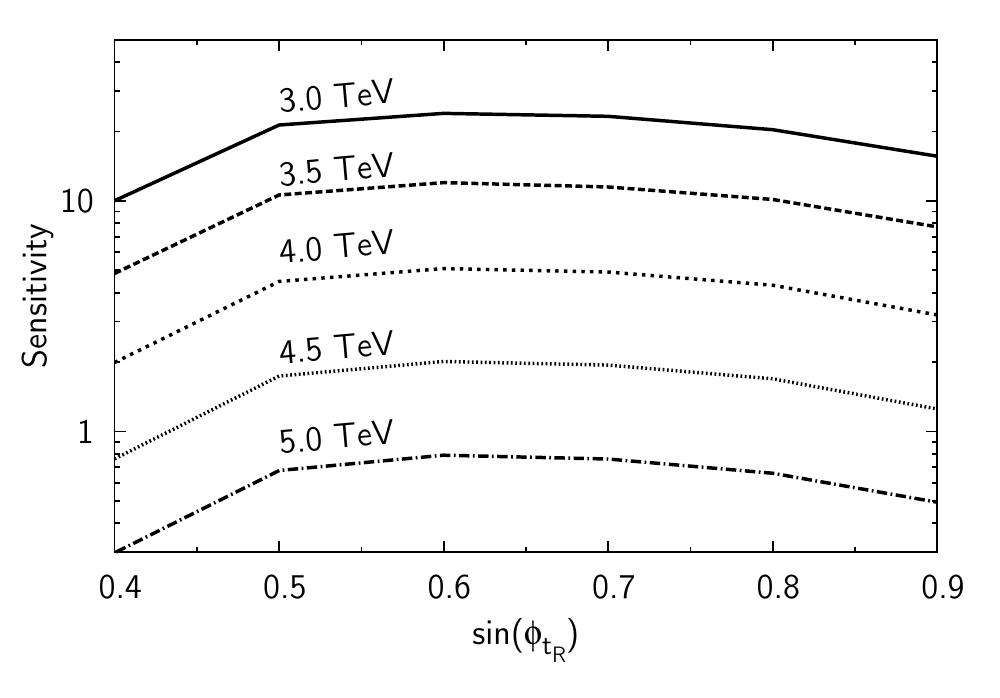}
\caption{Sensitivity obtained in the $Ht\bar{t}$ channel 
 for different values of the heavy gluon
  mass as a function of $\sin(\phi_{t_R})$ after an integrated
  luminosity of 100 fb$^{-1}$.
\label{Htt:results}
}
\end{minipage}
\hfil
\begin{minipage}{0.45\textwidth}
\includegraphics[width=0.95\textwidth]{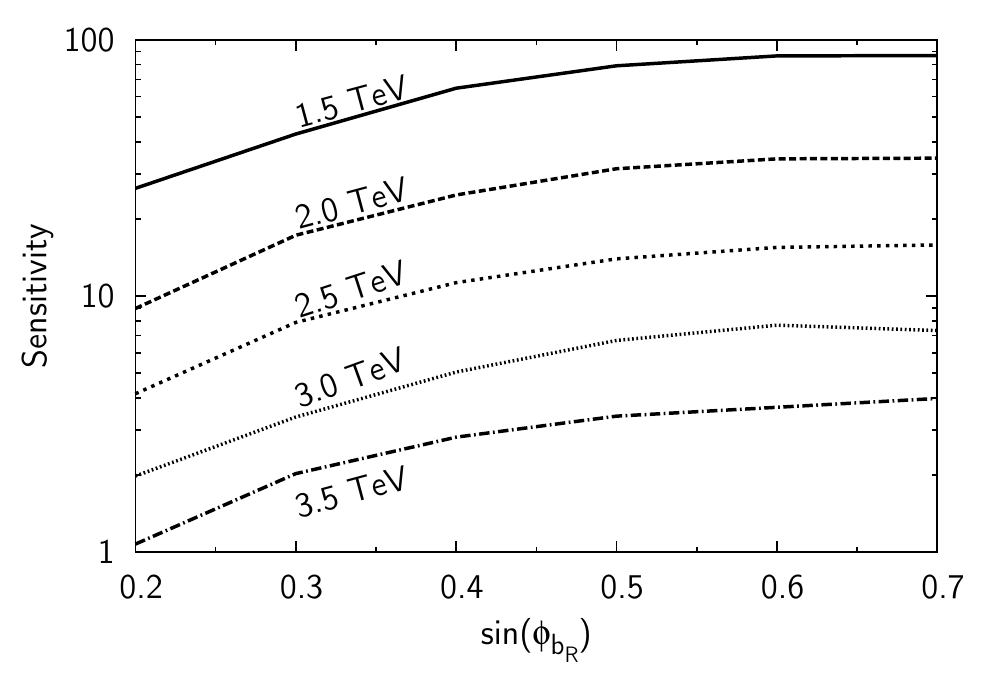}
\caption{Sensitivity obtained in the $Hb\bar{b}$ channel 
for different values of the heavy gluon
  mass as a function of $\sin(\phi_{b_R})$ after an integrated
  luminosity of 100 fb$^{-1}$.
\label{Hbb:results}
}
\end{minipage}
\end{figure}

\begin{figure}[h]
\hspace{0.5cm}
\includegraphics[width=0.45\textwidth]{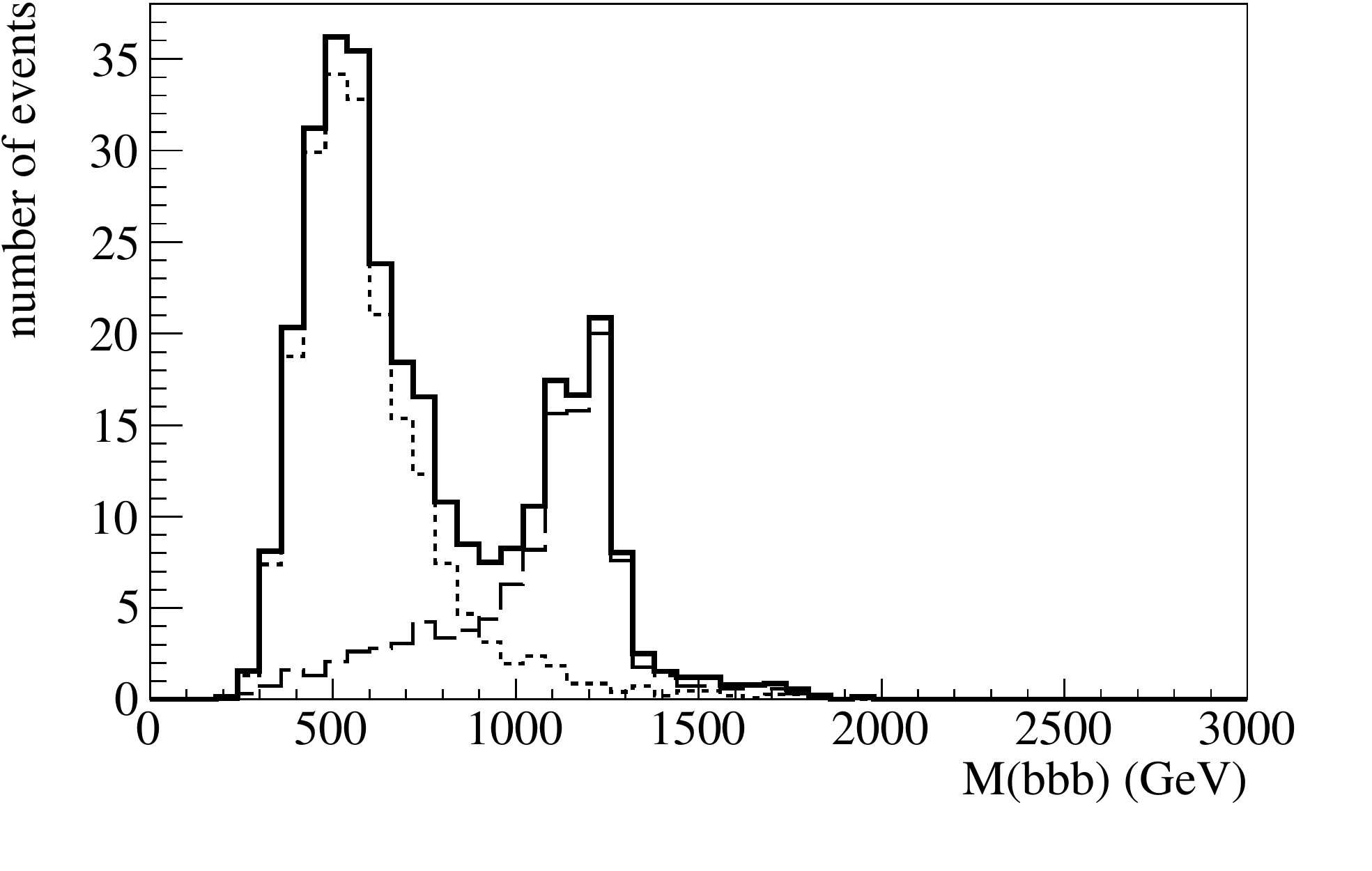}
\hfil
\includegraphics[width=0.45\textwidth]{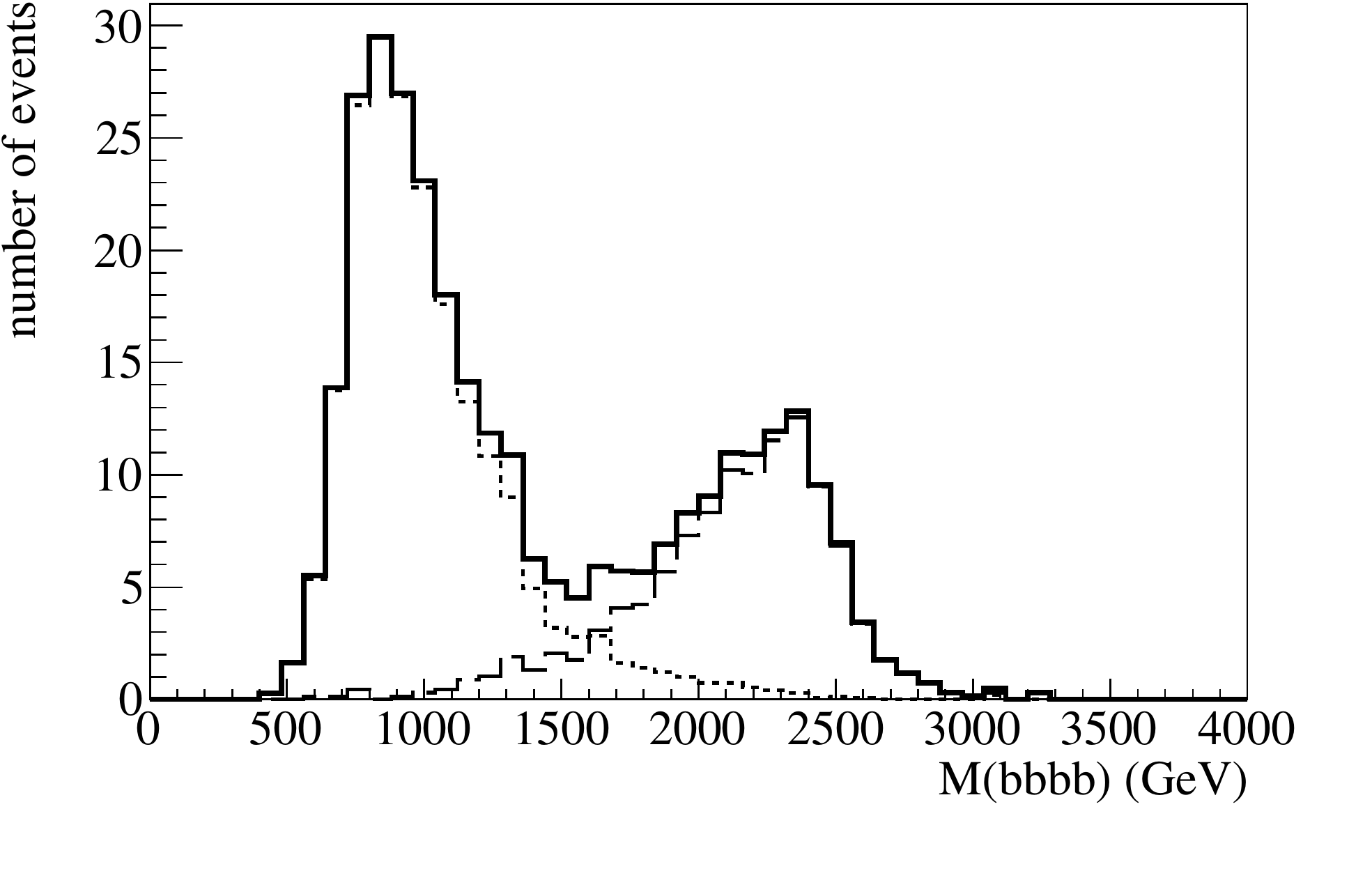}
\caption{Reconstruction of the heavy quark (left) and the heavy gluon
  (right) with $m_B=1250$ GeV and $m_G=2500$ GeV with an integrated
  luminosity of 100 fb$^{-1}$. 
The dotted, dashed and solid lines correspond to the background,
signal and sum of both, respectively.
\label{reconstruction}}
\end{figure}

In order to reduce the background to manageable levels we need to
require all four b-quarks to be measured, well isolated and also all
four of them to be quite hard. Once these cuts are imposed and the
signal and background cross sections are comparable, we can take
advantage of the particular kinematics of the signal in which the two
b-jets coming from the decay of the Higgs are typically softer than
the other two. Thus the requirement that the two softest b-jets
reconstruct the Higgs mass is a good extra discriminator. Finally,
we can reconstruct the $B$ quark using the reconstructed Higgs
together with the hardest b-jet and $G$ using the invariant mass of
all four b-jets. We use the latter invariant mass as a final discriminating
variable.  The proposed cuts are
\begin{itemize}
\item 4 b-tagged jets with $\Delta R(bb) > 0.7$ and $p_T(b)>50$ GeV
\item No isolated leptons
\item $p_T(b_h)>300$ GeV for the hardest b-jet
\item $|M_{b_3b_4} - m_H| < 30$ GeV, where $b_{3,4}$ are the two
  softest b-jets
\item $M_G-1000\mbox{ GeV} < M_{4b} < M_G+500$ GeV.
\end{itemize}
Our results are summarized in Fig.~\ref{Hbb:results} in which we show
the sensitivity that can be reached with 100 fb$^{-1}$ for different
values of $M_G$ as a function of $\sin(\phi_{b_R})$.
Further details of the study will be presented
elsewhere~\cite{inprep}. A sample of the reconstruction power for the
heavy quark and the heavy gluon is shown in Fig.~\ref{reconstruction}
for the particular case of $M_B=1.25$ TeV and $M_G=2.5$ TeV.

\section{Conclusions\label{conclusions}}

We have discussed how searches involving the Higgs boson can give us
valuable information on the top quark and other particles beyond the
SM related to the top quark. Current experimental data on the Higgs
boson might be pointing to a large correction to the top Yukawa
coupling, effectively reversing its sign. This possibility might be
tested through single top production in association with a Higgs boson
although further studies are required before considering the results
of such searches conclusive. We have also shown how searches involving
the production of $Ht\bar{t}$ and $H b\bar{b}$ can give us information
on the spectrum of new particles in models of strong EWSB. In
particular, they can be used to search for top partners, new vector-like
quarks that mix strongly with the top and/or bottom quarks. 

\ack
We would like to thank N. F. Castro and J. P. Araque for the useful discussions. This work has been supported by MICINN projects
FPA2006-05294 and FPA2010-17915, through the FPU programme and by Junta
de Andaluc\'{\i}a projects FQM 101, FQM 03048 and FQM 6552.

\end{document}